# On the use of Biplot analysis for multivariate bibliometric and scientific indicators

**Daniel Torres-Salinas**

EC3 Research Group: Evaluación de la Ciencia y la Comunicación Científica, Centro de Investigación Médica Aplicada,

Universidad de Navarra, 31008, Pamplona, Spain.

Email address: torressalinas@gmail.com

**Nicolás Robinson-García*, Evaristo Jiménez-Contreras**

EC3 Research Group: Evaluación de la Ciencia y la Comunicación Científica, Facultad de Comunicación y

Documentación, Colegio Máximo de Cartuja s/n, Universidad de Granada, 18071, Granada, Spain.

Telephone: +34 958 240920

Email addresses: elrobin@ugr.es; evaristo@ugr.es

**Francisco Herrera**

Departamento de Ciencias de la Computación e I. A., CITIC-UGR, Universidad de Granada, 18071, Granada, Spain

Email address: herrera@decsai.ugr.es

**Emilio Delgado López-Cózar**

EC3 Research Group: Evaluación de la Ciencia y la Comunicación Científica, Facultad de Comunicación y

Documentación, Colegio Máximo de Cartuja s/n, Universidad de Granada, 18071, Granada, Spain.

Telephone: +34 958 240920

Email addresses: edelgado@ugr.es

**Abstract**

Bibliometric mapping and visualization techniques represent one of the main pillars in the field of scientometrics. Traditionally, the main methodologies employed for representing data are Multi-Dimensional Scaling, Principal Component Analysis or Correspondence Analysis. In this paper we aim at presenting a visualization methodology known as Biplot analysis for representing bibliometric and science and technology indicators. A Biplot is a graphical representation of multivariate data, where the elements of a data matrix are represented according to dots and vectors associated with the rows and columns of the matrix. In this paper we explore the possibilities of applying the Biplot analysis in the research policy area. More specifically we will first describe and introduce the reader to this methodology and secondly, we will analyze its strengths and weaknesses through three different study cases: countries, universities and scientific fields. For this, we use a Biplot analysis known as JK-Biplot. Finally we compare the Biplot representation with other multivariate analysis techniques. We conclude that Biplot analysis could be a useful technique in scientometrics when studying multivariate data and an easy-to-read tool for research decision makers.

*To whom all correspondence should be addressed.

**1. Introduction**

Bibliometric mapping and visualization techniques represent one of the main pillars in the field of scientometrics. Nevertheless, Derek de Solla Price, considered as the father of scientometrics, already stated his wish to "exhibit an interlocking metabolic complex of bibliometric (and scientometric) parameters in a comprehensive and integrated structure after the manner of the Nitrogen Cycle" (Price as cited by Wouters, 1999). Since this statement, this research front has greatly expanded, especially in the seventies and eighties and was revitalized again in the late nineties due to technological advancements, as a tool for research policy monitoring (Noyons, 2001). The use of science maps has long been discussed in literature, emphasizing its capability as an easy-to-read tool that enables decision makers to understand the complexity and heterogeneity of scientific systems in order to rapidly respond to their behavior (Noyons & Calero-Medina, 2009).

Visualizing bibliometric data with scientific maps allows a better understanding of the relation between disciplines, invisible colleges or research fronts, for instance. According to Klavans & Boyack (2009), scientific maps can be defined as a two-dimensional representation of a set of elements and the relationship among them. Following this line of thought, for scientific mapping two techniques must be applied: firstly, a classification methodology, and secondly, a representation technique. Traditionally, the main classifying methodologies employed for representing bibliographic data have been those based on multivariate analysis such as Multi-Dimensional Scaling (MDS), Principal Component Analysis (PCA) or Correspondence Analysis, for instance. A review on the application of these methodologies for scientific mapping can be found in Börner, Chen & Boyack (2003). However, not many representation techniques have been used; focusing especially on Pathfinder Networks (PFNet) (White, 2003), Self-organizing maps (SOM) (Moya-Anegón, Herrero-Solana & Jiménez-Contreras, 2006) or social networks (Groh & Fuchs, 2011). Drawing a low-dimensional graph implies the loss of some of the information inherent not just to the represented elements, but also to the variables that affect their similarity or disimilarity.

Regarding these techniques, in this paper we aim at presenting a visualization methodology known as Biplot analysis (Gabriel, 1971) which could introduce interesting and useful novelties in scientific maps, opening new possibilities in the field of scientometrics. A Biplot is a graphical representation of multivariate data, where the elements of a data matrix are represented according to dots and vectors associated with the rows and columns of a matrix. Contrarily to a scatter gram, the axes are not perpendicular, as they simulate the projection of an n-dimensional representation over a surface with a minimum loss of information, adding interpretative meaning to the cosine of the angles between vectors, which represents the correlation between variables. Therefore, when vectors are perpendicular, the cosine equals zero and the variables are independent. But if they are very close or represent a 180º angle, they have a highly positive or negative correlation.

In short, the Biplot analysis is a graphical representation of multivariate data that mixes variables and cases (that is the reason for the bi prefix), enabling the user, to intuitively interpret for example in a bibliometric context; indicators and cases. Not as widely expanded as other techniques such as the above mentioned, it was first proposed by Gabriel (1971) and has already been tested in its many variants and types in very different scientific fields such as: Medicine (Gabriel, 1990), Genetics (Wouters et al, 2003), Agriculture (Yan et al, 2000), Library Science (Veiga de Cabo & Martín-Rodero, 2011), Economics and Business (Galindo, Vaz & Nijkamp, 2011), Tourism (Pan, Chon & Song, 2008) or Political Science (Alcántara & Rivas, 2007). Within the field of bibliometrics, this methodology was first introduced in conference paper in which the Biplot analysis was applied in order to analyze the scientific activity in Health Sciences of a small set of Spanish universities (Arias Díaz-Faes et al, 2011).

Considering the success and expansion the Biplot methodology has had in other research areas, the main objective of this paper is to deepen into the possibilities of applying the Biplot analysis in the field of scientometrics. More specifically, we aim at firstly describe and introduce this methodology to the reader and secondly, analyze its usefulness through three different case studies, showing its easy use for understanding and reading multivariate data in a research policy context. These case studies are chosen in order to explore the methodology's strengths and weaknesses when using different contexts, types of variables and levels of analysis. Then we use the first case study in order to compare this methodology with CA, MDS and PCA. The case studies proposed are the following:

- The first case study reflects the scientific efforts of European countries and their performance considering several bibliometric and S&T indicators.

- The second study will analyze the top 25 countries in the THE Ranking according to their performance in four of the variables it uses for ranking universities.

- Thirdly, we analyze a Spanish university's research performance in different research fields according to its output in the Thomson Reuters Web of Science databases.

This paper is structured as follows. In Section 2 we present and describe the classic Biplot methodology. Then, we describe three case studies, for which we will apply this representation method, for this, we select the JK-Biplot type. The results of these three cases along with a comparison with other multivariate techniques are shown and discussed in Section 3. In Section 4 we conclude with some remarks on the strengths and weaknesses of this technique. Appendix 'Biplot methodology in terms of spectral decomposition' has been included at the end of the paper in order to provide a more thorough description of the Biplot methodology.

## 2. Methodology

In this section we will present the Biplot analysis and briefly introduce three case studies in which we will apply it. This section is structured as follows. Firstly we give an overview on the Biplot analysis. In subsection 2.2, we give the key points for interpreting a Biplot representation and we introduce the JK-Biplot based on PCA, which is the one we will use for presenting the application of this methodology in the field of scientometrics. In subsection 2.3. we shortly introduce the software used for developing our applications. Then, in subsection 2.4., we introduce the three case studies used.

*2.1. A snapshot on the Biplot analysis*

As we have previously mentioned, Biplot is a data representation technique consisting on visualizing a matrix with more than two variables in a low dimensional graph where each row represents a subject and each column a variable. This technique is usually applied after a multivariate analysis has been performed, ranging from log-ratio analysis, principal component analysis or correspondence analysis; in fact to any method based on a singular-value decomposition. Due to its simplicity, its potentiality lies on enabling to visualize not just the relation between subjects or cases considering certain variables, but also the relationship between the variables.

Gabriel originally described three types of Biplot analysis, considered as the classical ones (Cárdenas et al, 2007) depending on the quality of representation of cases and variables. Therefore, we have: the GH Biplot Analysis, which emphasizes variables' representation, the JK Biplot Analysis, focused on the represented elements, and the SQRT Biplot Analysis, which tries to balance the quality of representation of the overall matrix. Other types of Biplot analysis are HJ Biplot analysis (Galindo, 1986) and GGE Biplot analysis (Yan et al, 2000).

The Biplot is based on the same principles as other factorial techniques for dimensionality reduction, with the only difference that in this case, it represents the data but also the variables, obtaining a dual representation between principal components and the main coordinates. Its interpretation is based upon geometric concepts which are intuitive for the user, facilitating their understanding. In Figure 1 the basic ideas for interpreting a Biplot representation are explained:

- The similarity of subjects (rows) is the inverse function of the distance between them.

- The length and angles of the vectors (columns) represent variance and covariance respectively.

- The relation between rows and columns must be understood as dots products, that is, the projection of the cases over the variables.

FIGURE 1. Basic interpretation of a Biplot representation

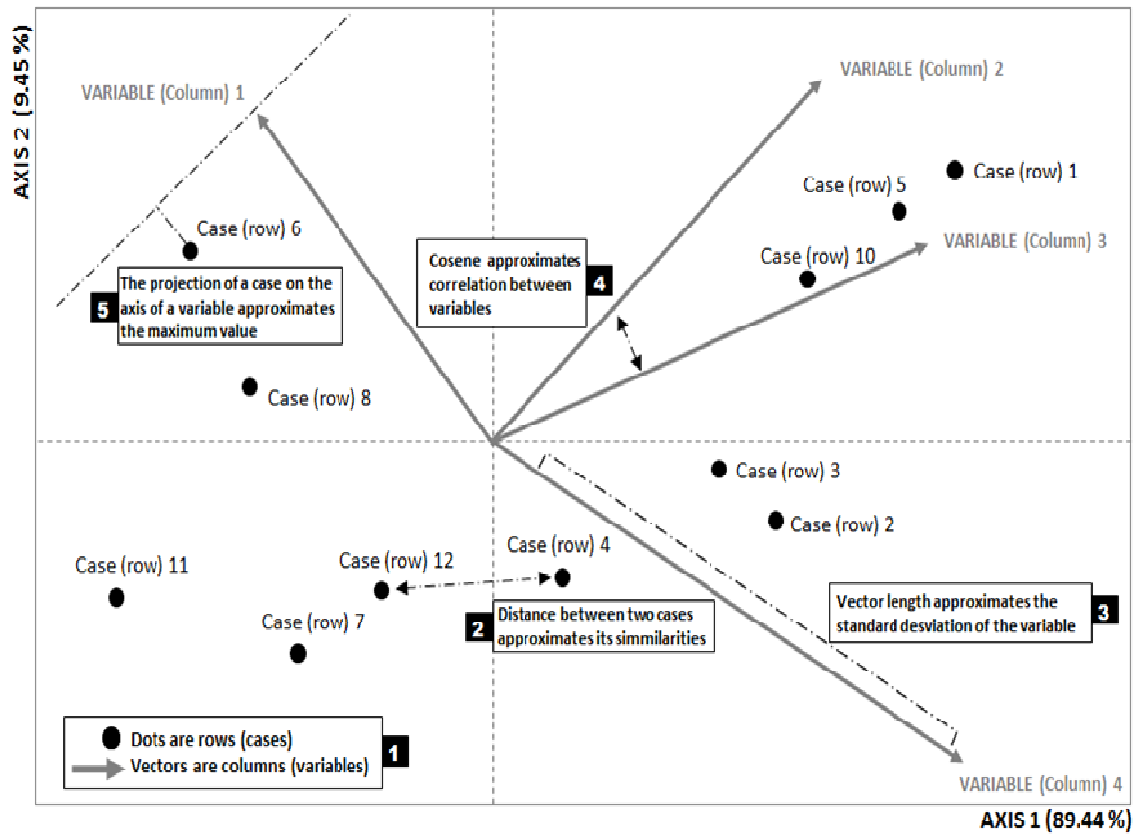

Following this Figure we shortly introduce the 5 elements to take into consideration in the future analysis:

1. Dots are rows (cases) and vectors are columns (variables).

2. The distance between two cases approximates its similarity.

3. The vector length approximates the standard deviation of the variables.

4. The cosine between two vectors approximates the correlation between variables.

5. The projection of a case on the axis of a variable approximates the maximum value.

*2.2. Biplot methodology*

A Biplot is defined as a low-dimensional graph with a minimum loss of information of a given matrix of data $X_{(n \times p)}$, formed by markers $a_1, a_2, ..., a_n$ for rows and $b_1, b_2, ..., b_p$ for columns, chosen in such a way that each element $x_{ij}$, is an approximation to $x_{ij} = a_i^T b_j$ (Gabriel, 1971). In this subsection we will focus on providing clear rules for interpreting a Biplot representation. For a more exhaustive presentation of this methodology in terms of spectral decomposition, the reader is referred to Appendix Biplot methodology in terms of spectral decomposition.

The Biplot methodology offers approximate representations in a plane for data matrices with more than two dimensions that would otherwise, have to be represented in n-dimensions being n the number of variables. Variables are represented by linear axis with scales in the same way as in a normal scatter gram. Markers are located by projecting their mark perpendicularly onto the axes for variables (columns) and reading the value on the scale. These projected scale values are approximations of the true values as it is not usually possible to represent more than two variables exactly in the plane. Coordinates of markers are obtained from a PCA or a CA for instance, where the position of a marker is defined by the first two principal components. Also, the coordinates of variables are obtained with respect to the first two principal components, each weighted by the standard deviation of that component that is by the square root of the corresponding eigenvalue.

As observed in figure 1, any two correlated variables are represented with their biplot axes pointing to similar directions, as markers with a high or low value for one of the correlated variables will have similar values for the other variable. On the contrary, if variables are correlated negatively, markers with a high value for one of the variables will presumably have a low value for the other variable. This means that correlation between variables can be obtained from the angle they form. Therefore, an acute angle between variables will presume a positive correlation among them; an obtuse angle will presume a negative correlation; and a right angle, no correlation between variables. These correlations are approximately represented by means of the cosines of the angles.

Another important aspect when interpreting a Biplot representation has to do with the display of the axes. Normally, these meet at the centroid which is the mark for the means of all the variables. Also, the length of the vectors (variables) is significant, as it displays the approximate value of the standard deviation of the variables. Depending on the preservation of columns or rows during the factorization we may have a Row Metric Preserving (RMP) Biplot or a Column Metric Preserving (CMP) Biplot. This two types are called a JK-Biplot and a GH-Biplot respectively and their main differences have to do with their emphasis for better representing rows than columns (JK-Biplot) and viceversa (GH-Biplot). In order to produce a symmetric Biplot we would need to balance the preservation values for columns and rows, this is what is called a SQRT Biplot.

In this paper we will use the JK-Biplot in order to explore its possibilities as it is the most common type. Its main feature is that the scalar product of the markers reproduces the matrix element. This concept is fundamental to geometrical interpretation in terms of distances, angles, orthogonal, etc.

Let consider a given set of data where the markers for rows and columns in a s dimension are:

$$A_{(s)} = J_{(s)} = U_{(s)}\Lambda_{(2)} B_{(s)} = K_{(s)} = V_{(s)}$$

This variant of Biplot analysis presents the following advantages.

Firstly, dot products with identical metric from rows of matrix X, coincide with the dot products of markers contained in J. The approximation of these dot products in a low-dimensional graph is optimal considering their minimum squares. In fact:

$$XX' = JK'KJ' = JJ'$$

Also, the spectral decomposition of the dot products matrix between rows is also the decomposition of its singular values:

$$XX' = U\Lambda^2 U$$

then, the best approximation to range s is:

$$XX' = U_{(s)}\Lambda^2_{(s)}U'_{(s)} = J_{(s)}J'_{(s)}$$

which coincides with the one obtained in the Biplot of matrix X.

Consequently, the Euclidean distance between two rows of X coincide with the Euclidean distance between markers J.

Also, markers for rows coincide with the coordinates for each case in a principal components space:

$$XV_{(s)} = U\Lambda\,V'V_{(s)} = U_{(s)}\Lambda_{(s)} = J_{(s)}$$

This means we can study similarities between cases with a minimum information loss.

Secondly, markers for rows coincide with the coordinates assigned to each case in the principal component space. In order to demonstrate this property, let consider V a matrix containing vectors from S, then coordinates over the first *s* components can be described as:

$$XV_s = (UDV')V_s = U_s D_s = J_s$$

This means that, when the Euclidean distance is adequate for the analysis, one can study similarities among the cases according to their markers.

Thirdly, the coordinates for columns are projections over the original axes in the principal components space. That is, coordinates of the vectors that construct the canonical base can be described as an identity matrix $I_p$ and the projection of these over the principal components spaces can be described as:

$$I_p V_{(s)} = V_{(s)} = K_{(s)}$$

This means that coordinates for columns fix the unit for prediction scales. This property allows interpreting coordinates as the correlation between the original variables and the axes.

Finally, the last property of the JK-Biplot has to do with the quality of the representation. As mentioned above, this type of Biplot represents better rows than columns, contrarily to the GH-Biplot which emphasizes columns over rows.

*2.3. 'MultBiplot' Software*

For this study we have used the free beta version of the software 'MultBiplot' developed by Vicente-Villardón (http://Biplot.usal.es/multBiplot). This program implements the experience of the 'Applied Statistics Group' at the University of Salamanca (Spain) in working on Biplot analysis. According its authors this software is conceived not to be "another Biplot program", but to fill the gap between the static pictures and a more dynamic visual interpretation. So it is specialized on improving the visualization of Biplot diagrams. In relation to the different Biplot techniques, this program contains the Classical Biplot (JK) as well as the HJ-Biplot proposed by Galindo (1986). From the users' viewpoint, the 'MultBiplot' software does not require any kind of special training or a long learning period, being highly recommended for those who want to learn this statistical technique.

### 2.4. Data source and indicators

TABLE 1. Description of the indicators used in the three different study cases

| Indicator / Measure | Definition* | Acronym | Source |
|---|---|---|---|
| **CASE 1: Countries** | | | |
| Share of human resources in S&T | Labor force working in S&T from the total share of a country | %HR | Eurostat |
| R&D expenditure (Millions of €) | Total budget of countries devoted to R&D activities | MILL € | Eurostat |
| R&D expenditure (Percentage of GDP) | Proportion of countries' Gross Domestic Product devoted to R&D activities | GDP | Eurostat |
| Total Researchers | Total number of professionals devoted to activities related with R&D | RES | Eurostat |
| Number of Citations | Total number of citations received by publications generated by each country according to the Scopus database | CIT | SJ&CR |
| Number of Citable Documents | Citable documents are considered those published by journals indexed in Scopus under the following document types: articles, reviews and conference papers | DOC | SJ&CR |
| Citation Average | Average of citations received per citable document | CAVG | SJ&CR |
| Normalized Citation Average | Ratio between the average scientific impact of an institution and the world average impact of publications | NCIT | SJ&CR |
| **CASE 2: Universities** | | | |
| Research | Volume, income and reputation | RESEARCH | THE Ranking |
| Citation | Research influence | CITATION | THE Ranking |
| International Outlook | Staff, students and research | INT OUTLOOK | THE Ranking |
| Teaching | Learning environment | TEACHING | THE Ranking |
| **CASE 3: Scientific Fields** | | | |
| Citation Average | Average of citations received per document | ACIT | Thomson Reuters |
| Percentage of Top Cited Papers | Share of the total output of a university included in the top 10% of the most highly cited documents in the field according to the national output | TOPCIT | Thomson Reuters |
| Percentage of Fist Quartile Papers | Share of documents published in journals ranked in the top 25% according to the Thomson Reuters Journal Citation Reports | %Q1 | Thomson Reuters |
| Number of Citations | Total number of citations received by documents published by a university in a given field | NCIT | Thomson Reuters |
| H-Index (Hirsch) | Number of documents (h) published by a university in a given set that has received at least h citations | H-Index | Thomson Reuters |
| Number of Citable Documents | Citable documents are considered those published by journals indexed in Thomson Reuters Web of Science under the following document types: articles, reviews notes and letters | NDOC | Thomson Reuters |

* Definitions for variables in case 2 are displayed as stated in http://www.timeshighereducation.co.uk/story.asp?storycode=417368

Considering that the aim is to present the Biplot analysis representation technique, three basic study cases were chosen, representing three different research evaluation contexts. Although this technique is usually applied to large data collections, in this paper we chose cases with a smaller size in order to ease the interpretation of the representation to the reader. We selected the JK-Biplot type which emphasizes cases representation over variables, and we used Principal Component Analysis as a classification methodology and data reduction. The three cases selected were: scientific effort and bibliometrics indicators of European countries, top Universities in the THE Ranking and the University of Granada's research performance in 12 different scientific fields. The selected data sources and the variables for each case study are displayed in Table 1. For more specific data regarding goodness of fitness and quality of representation ($QR_{overall}$, $QR_{col}$ and $QR_{row}$) for each case, the reader is referred to http://www.ugr.es/~elrobin/QR_On_the_use_of_Biplot.xlsx where an excel file can be obtained with all the details.

**3. Analysis and results**

In the following 3 subsections we present the analysis and results for each case study. Finally, we briefly compare the results of one of the study cases with those given by applying other techniques (PCA, MDS, CA) in order to show the advantages of the Biplot representation in comparison with other methodologies for interpreting multivariate data with more than two variables. Usually, these techniques join together the information given by the variables, introducing two artificial variables instead and therefore, losing some information in the representation.

*3.1. Case 1. Scientific effort and bibliometrics indicators for European Countries*

We analyze the research performance and input of a set of European countries. For this analysis we considered a 21x8 matrix where rows correspond to European countries and columns to indicators regarding R&D efforts and bibliometric indicators. The study time period used was 2009 or 2010. Data regarding R&D indicators was extracted from the EUROSTAT Portal, while bibliometric indicators were extracted and calculated from data retrieved from the Scimago Journal & Countries Rank databases. Countries and indicators are presented in table 2.

TABLE 2. Science & Bibliometrics for European Countries

|  | MILL € | GDP | RES | %HR | DOC | CIT | CAVG | NCIT |
|---|---|---|---|---|---|---|---|---|
| Germany | 69810 | 2.82 | 484566 | 44.8 | 119216 | 228773 | 1.76 | 1.36 |
| France | 43633 | 2.26 | 295696 | 43.9 | 87430 | 148995 | 1.57 | 1.39 |
| United Kingdom | 30071 | 1.77 | 385489 | 45.1 | 123756 | 253482 | 1.81 | 1.42 |
| Italy | 19539 | 1.26 | 149314 | 33.8 | 67459 | 118043 | 1.6 | 1.23 |
| Spain | 14588 | 1.39 | 221314 | 39 | 59642 | 96368 | 1.48 | 1.10 |
| Sweden | 11869 | 3.42 | 72692 | 50.8 | 25257 | 54567 | 2.03 | 1.39 |
| Netherlands | 10769 | 1.83 | 54505 | 51.9 | 39499 | 96134 | 2.22 | 1.66 |
| Austria | 7890 | 2.76 | 59341 | 39.2 | 15476 | 31879 | 1.9 | 1.23 |
| Denmark | 7208 | 3.06 | 52568 | 51.9 | 15042 | 38504 | 2.38 | 1.60 |
| Belgium | 7047 | 1.99 | 55858 | 49.3 | 21978 | 46169 | 1.95 | 1.44 |
| Finland | 6971 | 3.87 | 55797 | 50.6 | 13308 | 25310 | 1.81 | 1.26 |
| Norway | 5342 | 1.71 | 44762 | 51.5 | 12755 | 22401 | 1.62 | 1.39 |
| Ireland | 2796 | 1.79 | 21393 | 45.9 | 9499 | 17728 | 1.73 | 1.24 |
| Portugal | 2747 | 1.59 | 86369 | 23.9 | 12957 | 16756 | 1.22 | 1.05 |
| Poland | 2607 | 0.74 | 98165 | 36.3 | 26057 | 23729 | 0.88 | 0.64 |
| Czech Republic | 2334 | 1.56 | 43092 | 37.8 | 13790 | 17005 | 1.18 | 0.77 |
| Hungary | 1126 | 1.16 | 35267 | 33 | 7542 | 10648 | 1.34 | 0.91 |
| Slovenia | 745 | 2.11 | 10444 | 40.8 | 4104 | 4697 | 1.1 | 1.05 |
| Romania | 572 | 0.47 | 30645 | 24.4 | 10897 | 6254 | 0.56 | 0.73 |
| Slovakia | 416 | 0.63 | 21832 | 33.5 | 4195 | 4043 | 0.93 | 0.72 |
| Bulgaria | 214 | 0.6 | 14699 | 31.6 | 3293 | 2285 | 0.68 | 0.74 |

In Figure 2 we show the Biplot representation of this case. The goodness of fit is 89.9%. All variables (columns) are well represented as they all have a $QR_{col}$ above 0.95 except GDP where it reaches 0.75. Rows are also well represented, 15 countries present a $QR_{rows}$ above 0.90 and 6 between 0.73 and 0.86. Regarding the variables two latent variables can be clearly distinguished in the graph, indicating a high correlation between the observed variables of each of them. Therefore, the correlation between %HR and DOC is 0.198 and between CAVG and NCIT is 0.928. The first latent variables which encompasses Human resources (%HR), %GDP, average of citations (CAVG) and normalized citations (NCIT) could be defined as the qualitative axis as these measures are all normalized. The second latent variable, which is formed by variables related with raw indicators influenced by size (CIT, MILL €, DOC, RES) could be defined as one of a quantitative measure.

FIGURE 2. JK-Biplot analysis for European Countries according to their Science & Bibliometric Indicators

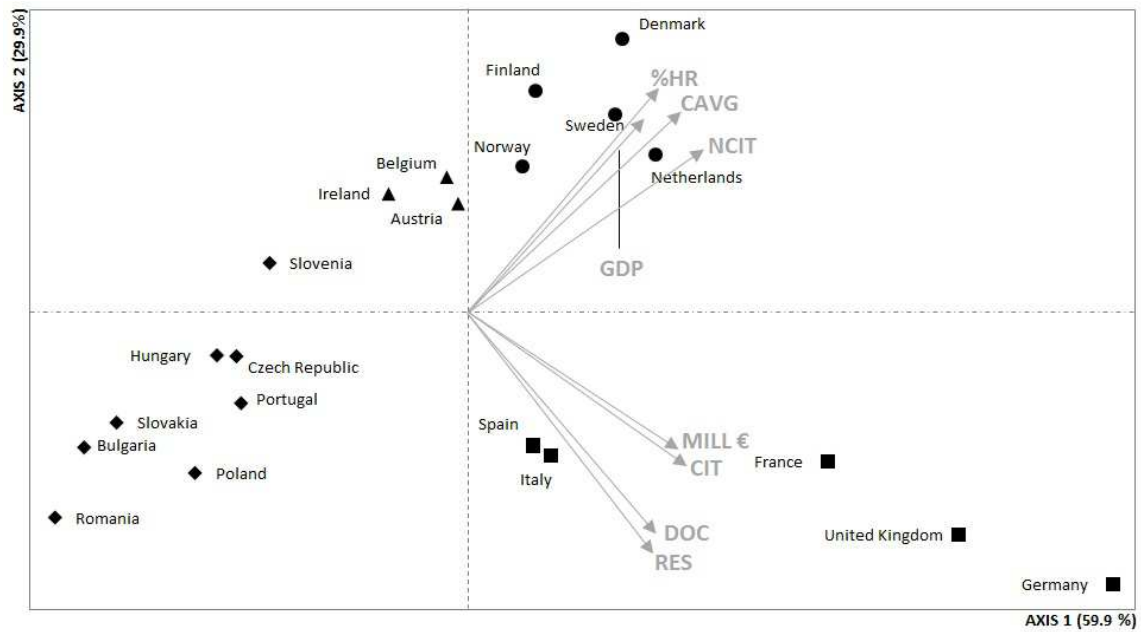

In regard to the countries, we observe four distinct groups according to their scientific profile.

- There is a group formed by the Nordic countries (Norway, Sweden, Denmark and Finland) and the Netherlands (upper right), characterized as big investors in science (%HR and GDP) and with a high scientific impact (CAVG and NCIT).

- A second cluster can be observed (lower right) where countries such as Germany and United Kingdom and France perform well in all variables; effort and bibliometric indicators. A subset of this second group is formed for two Mediterranean countries; Spain and Italy, with lower values for normalized bibliometric indicators and less R&D efforts than the other members of this cluster and the first one.

- Another cluster can be found (upper left) formed by four small countries (Belgium, Ireland, Austria and Slovenia) characterized for a medium performance regarding R&D efforts and bibliometric indicators.

- Finally, we find countries (lower left), - mainly from east Europe as Bulgaria, Romania, Hungary, etc. - characterized by their low investment on R&D and their low research performance.

Consequently, we observe how this representation allows the reader to easily spot countries that are similar, not just regarding to their geographical location, but also to their scientific culture.

*3.2. Case 2.Top Universities in the THE Ranking*

We analyze 'world-class universities' performance according to the variables used in the Times Higher Education World University Ranking. We considered a 25x4 matrix where rows correspond to the top 25 universities from the 2012 and columns correspond to the different indicators and measures employed in this classification. That is: Teaching, Research, Citations and International Outlook. Industry Income was excluded for this analysis as data is not provided for all universities. A more thorough description of the methodology employed by this ranking is available at THE rankings website. Values for each university and variable are shown in Table 3. Figure 3 shows the Biplot representation.

TABLE 3. Top 25 universities according to the THE Ranking variables (data: 2012 edition)

|  | Teaching | International Outlook | Research | Citations |
|---|---|---|---|---|
| ETH Zürich - | 79.1 | 97.5 | 85.8 | 87.2 |
| Imperial College London | 88.8 | 92.2 | 88.7 | 93.9 |
| University of Oxford | 89.5 | 91.9 | 96.6 | 97.9 |
| University College London | 77.8 | 91.8 | 84.3 | 89 |
| University of British Columbia | 68.6 | 88.7 | 78.6 | 85.2 |
| University of Cambridge | 90.5 | 85.3 | 94.2 | 97.3 |
| Massachusetts Institute of Technology | 92.7 | 79.2 | 87.4 | 100 |
| University of Toronto | 76.9 | 69 | 87.4 | 86.5 |
| Columbia University | 89.1 | 67.6 | 81.8 | 97.8 |
| Harvard University | 95.8 | 67.5 | 97.4 | 99.8 |
| Georgia Institute of Technology | 66.6 | 65 | 73.8 | 91.9 |
| Johns Hopkins University | 78.9 | 59.9 | 86.5 | 97.3 |
| University of Chicago | 89.4 | 58.8 | 90.8 | 99.4 |
| Stanford University | 94.8 | 57.2 | 98.9 | 99.8 |
| California Institute of Technology | 95.7 | 56 | 98.2 | 99.9 |
| Yale University | 92.3 | 55.5 | 91.2 | 96.7 |
| Carnegie Mellon University | 65.7 | 55 | 79.5 | 97.4 |
| CornellUniversity | 70.4 | 53.4 | 87.2 | 93.5 |
| University of California Berkeley | 82.8 | 50.4 | 99.4 | 99.4 |
| Princeton University | 91.5 | 49.6 | 99.1 | 100 |
| University of Michigan | 75.4 | 47.2 | 90 | 94.3 |
| Duke University | 62.6 | 46.9 | 77.9 | 97.4 |
| University of California Los Angeles | 85.9 | 41 | 92.5 | 97.3 |
| University of Washington | 70.8 | 36.9 | 74 | 98.2 |
| University of Pennsylvania | 87 | 34.3 | 86.1 | 97.9 |

The goodness of fit is 87.9%. Rows are represented with a $QR_{row}$ above 90% for 17 universities, 80% for 3 universities and less than 75% for 5 universities. Michigan, MIT and Columbia have the lower $QR_{row}$ as they have most of the information represented in axis 3 which is the one not covered in our biplot representation. In regard to columns, their $QR_{col}$ is above 80% for all variables. When observing the overall representation, we must point out that, firstly, two variables do not correlate with the rest (Citations and International Outlook) and secondly, two other variables are very closely related to each other (Research and Teaching). In this last case the correlation value is 0.784. Regarding to the cases, there are four distinct clusters of universities.

   - The first cluster (lower right) is formed by the universities with the highest values on Teaching and Research and which display a good performance in Citations. For instance, we see the two top British universities along with different universities from the North-American Ivy League such Harvard or Yale, and universities from the West-Coast such California Berkeley or Caltech.

   - Secondly, we find those universities which perform better in Citations but which are not in top positions in Teaching and Research, such as Pennsylvania and California Los Angeles.

   - The third group (upper left) are universities that display the lowest performance in all indicators, such as Duke, Cornell or Michigan. This last group also coincides with the last top 25 universities in the THE Ranking.

   - Finally the last group (lower left) is the one formed by those universities characterized mainly by their high values in International Outlook but not in the other indicators. We can distinct in this cluster the main universities from London (University College and Imperial College) and also from Canada (Toronto and British Columbia).

FIGURE 3. JK-Biplot Analysis for top 25 universities according to the THE Ranking

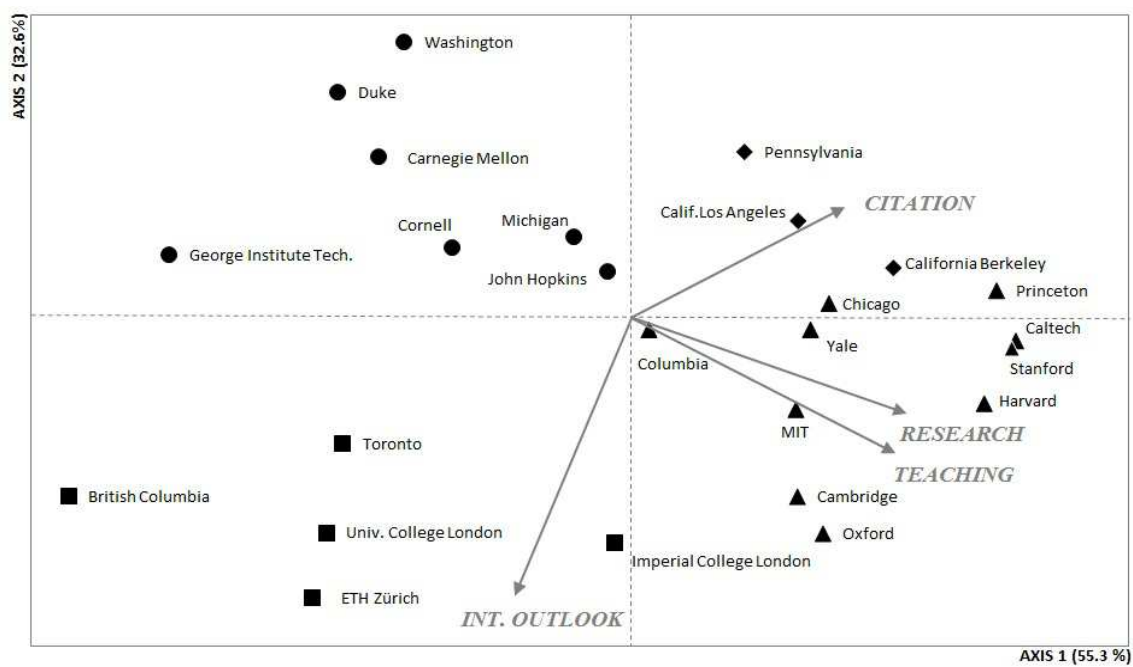

*3.3. Case 3. Scientific performance of the University of Granada in 12 scientific fields*

We analyze a university's research performance in 12 different scientific fields. For this, we selected the University of Granada (Spain) as a case study. We considered a 12x6 matrix where each row represents a scientific field and each column a bibliometric indicator regarding production and impact. Indicators were normalized according to all Spanish universities, meaning that the university with the best performance for a given indicator would reach a score of 1.00. We used the Thomson Reuters Web of Science databases and we selected 2006-2010 as the study time period. For more information over this data set, the reader is referred to Torres-Salinas et al. (2011a) and Torres-Salinas et al. (2011b). Indicators for each field of endeavor are shown in Table 4. In Figure 4 we illustrate the Biplot representation of this study case.

TABLE 4. Bibliometrics Indicators of the University of Granada in 12 Scientific Fields

| | Bibliometrics Indicators | | | | | | Normalized Indicators | | | | | |
|---|---|---|---|---|---|---|---|---|---|---|---|---|
| | NDOC | NCIT | H-Index | %1Q | ACIT | TOPCIT | NDOC | NCIT | H-Index | %Q1 | ACIT | TOPCIT |
| Agricultural Sciences | 174 | 821 | 14 | 72% | 4.71 | 17% | 0.352 | 0.408 | 0.737 | 0.885 | 0.854 | 0.733 |
| Biological Sciences | 958 | 5575 | 28 | 38% | 5.81 | 8% | 0.329 | 0.244 | 0.622 | 0.548 | 0.543 | 0.385 |
| Earth Sciences | 993 | 4567 | 23 | 54% | 4.59 | 11% | 0.729 | 0.577 | 0.742 | 0.891 | 0.658 | 0.579 |
| Economics & Business | 103 | 255 | 8 | 14% | 2.46 | 18% | 0.350 | 0.300 | 0.571 | 0.275 | 0.677 | 0.961 |
| Physics | 834 | 11763 | 28 | 62% | 14.1 | 11% | 0.374 | 0.577 | 0.560 | 0.793 | 1.000 | 0.662 |
| Engineering | 630 | 2699 | 22 | 61% | 4.28 | 12% | 0.320 | 0.381 | 0.733 | 0.844 | 0.465 | 0.643 |
| Mathematics | 777 | 1964 | 16 | 37% | 2.52 | 10% | 0.860 | 0.798 | 0.762 | 0.638 | 0.525 | 0.523 |
| Medicine & Pharmacy | 1412 | 8496 | 33 | 39% | 6.01 | 10% | 0.270 | 0.171 | 0.452 | 0.653 | 0.628 | 0.650 |
| Social Sciences | 263 | 503 | 11 | 30% | 1.91 | 9% | 0.809 | 0.652 | 0.917 | 0.523 | 0.584 | 0.315 |
| Psychology | 448 | 1477 | 16 | 23% | 3.29 | 12% | 0.911 | 0.652 | 0.800 | 0.376 | 0.456 | 0.335 |
| Chemistry | 1006 | 5595 | 26 | 58% | 5.56 | 8% | 0.376 | 0.262 | 0.591 | 0.813 | 0.534 | 0.379 |
| Inf. Technology | 502 | 2205 | 20 | 34% | 4.39 | 19% | 0.584 | 1.000 | 1.000 | 0.689 | 0.891 | 0.942 |

In this third case the goodness of fit is 72.2 %. It is the lower of three study cases presented. The $QR_{row}$ is over 80% in 8 scientific fields but it is insufficient in one of the other three; Economics & Business where it is 47%. In this field, most of the information is represented the third axis, however, no variables are represented there. Therefore, no conclusion can be obtained for this field after interpreting Figure 4. A similar situation occurs with columns where the $QR_{col}$ in five variables has a fit over 95% but one, %1Q, which is not well represented in axes 1 and 2. %1Q has a $QR_{col}$ of 3%. Relating with the representation, we observe that variables/vectors are grouped into clusters according to their correlation. On the left side we find relative variables such as Top Cited Documents (TOPCIT) and Citation Average (ACIT) which are size independent. On the right side we find such as Number of Citations (NCIT), H-Index and Citable Documents (NDOC) which are related to the raw data. We find the highest correlation values between NCIT and H-Index with 0.822 and the lowest between H-Index and TOPCIT with a correlation value of -0.042.

FIGURE 4. Biplot analysis of the University of Granada in 12 scientific fields according to bibliometric indicators

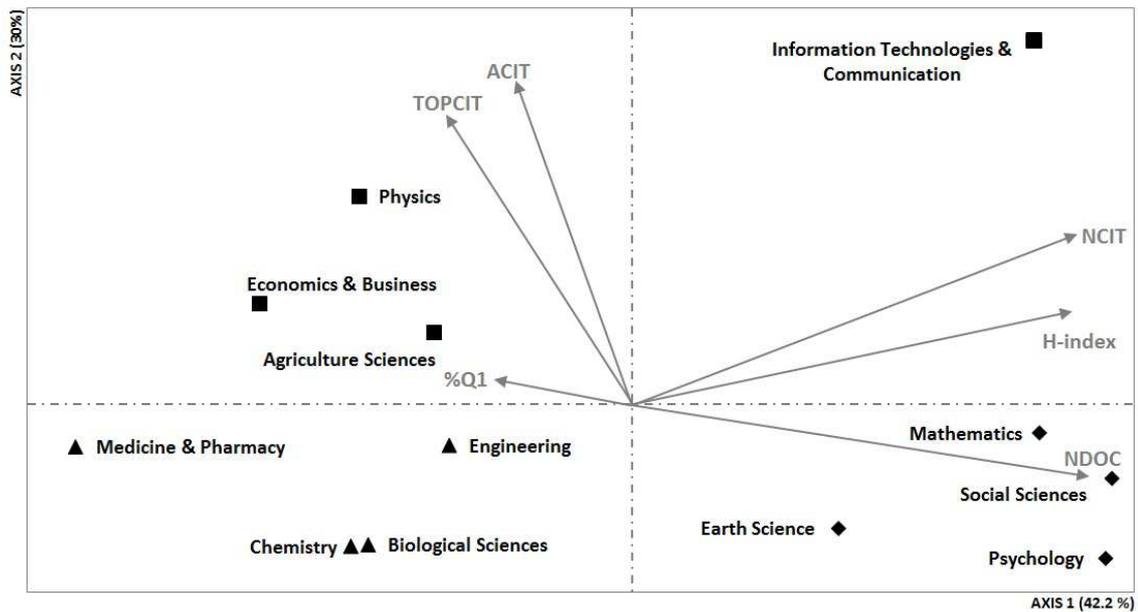

When observing the University of Granada's behavior regarding each scientific field (cases), we must outline the following:

- Two latent variables emerge from the observed variables. As in case 1, we have on the one hand the qualitative axis formed by TOPCIT, ACIT and %Q1 and a quantitative axis formed by NCIT, H-index and NDOC.

- It is highly significant the position of the Information Technology & Communication field (upper right) which stands completely by itself and separate from the rest of the fields. This is due to the high values it has for indicators of both latent variables except for %Q1.

- On the lower right side we find those fields on which the University of Granada outstands at national and internal level for raw indicators such as NDOC, H-Index or NCIT, that is for the quantitative axis. For example the University Granada is the second and third most productive university in Mathematics and Earth Sciences respectively in Spain, explaining its high values for variable NDOC.

- On the upper left side we find those areas in which the university performs well for qualitative indicators. In this sense, we must emphasize Physics and Agricultural Science for two indicators; TOPCIT and ACIT. In the case of Physics, it shows the best performance for TOPCIT of all fields, as reflected in the biplot. We also find Economics along with the %Q1 variable which had been previously discussed and cannot be interpreted in this representation due to the lack of information.

- Finally, we find a fourth group of areas in which this University of Granada has the worst performance according to the indicators displayed, for instance, Chemistry or Engineering. In fact these fields are where Granada is positioned lower in national rankings.

*3.4. Comparing JK Biplot representation with other multidimensional representation techniques*

Finally, in Figure 5 we present different visualization techniques applied to the first study case. Along with a JK Biplot representation we apply Correspondence Analysis (CA), Multidimensional Scaling (MDS) and Principal Component Analysis (PCA). We have chosen these techniques as they are the most common ones used for representing data in the field of bibliometrics. PCA is a mathematical methodology that uses orthogonal transformation converting a set of cases of possibly correlated variables in a set of values of uncorrelated variables which are known as principal components aiming at reducing the number of variables and guaranteeing that these are independent when data is jointly normally distributed. CA is a multivariate statistical methodology similar to PCA, providing the means to display and summarize a set of data in a two-dimensional graph. MDS is a visualization technique used for exploring similarities and dissimilarities in data. In the case of PCA and MDS we used the statistical software SPSS version 20.00. In the case of CA we used the statistical package XLSTAT and we used the Correspondence Factor Analysis with symmetric distances.

When comparing with MDS and PCA, Biplot representation offers a better solution, as the former are incapable of representing both, variables and cases, at the same time. However, even if it is done separately, MDS and PCA representations show similar patterns to those presented by the Biplot representation; with countries grouped in a similar way. For instance, the Biplot map and the MDS map show a very similar display of countries. Also, the PCA representation shows a similar pattern. In fact, the left corresponds with the lower right of MDS and Biplot with Germany and the UK outstanding, followed by France. The Nordic countries are displayed closely to each other as well as the pair Italy and Spain.

FIGURE 5. Representation of the case 1 (countries) using different multivariate techniques

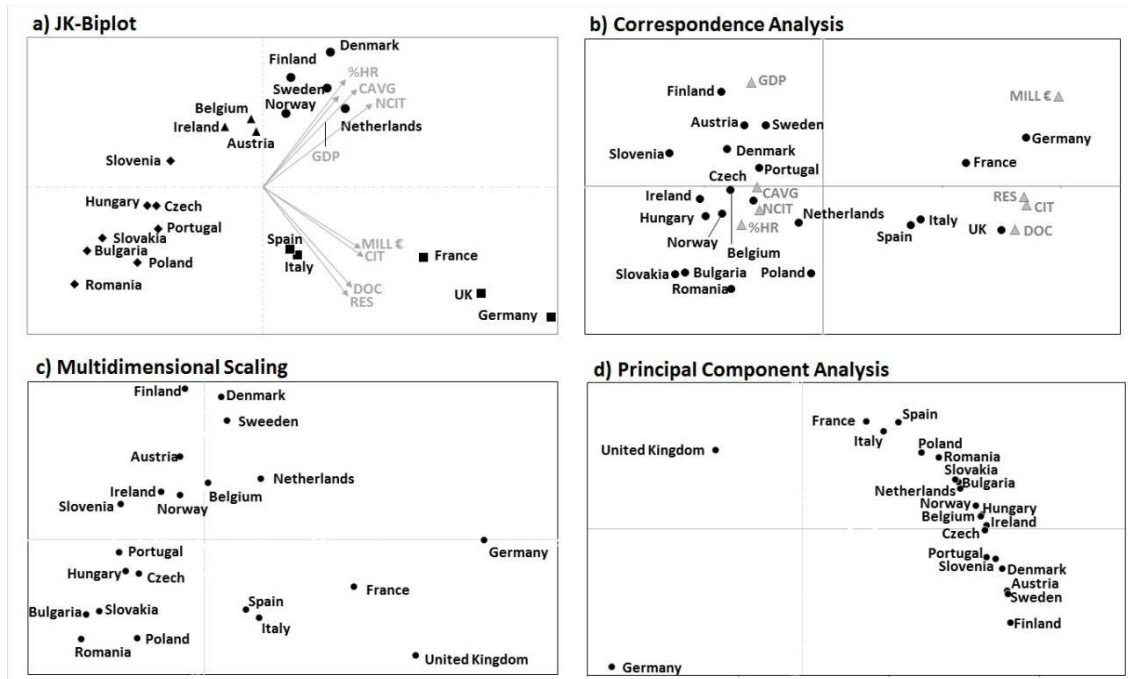

But if there is a method similar to the Biplot technique, that is the Correspondence Analysis (CA). This technique also represents rows and columns of a matrix, i.e. a contingency matrix, in a bidimensional graph. However, although the CA representation displayed in Figure 5 is similar to the Biplot map, we find it much more difficult to interpret as the relation between variables and cases is not perceived as easily as it occurs with the Biplot representation. Also, as it happened with the other two techniques, it offers a poorer representation losing much of the information, especially regarding the visualization of variables where the Biplot analysis displays their correlation between each other and their standard deviation. For these reasons many authors (Gabriel, 2002) point out the Biplot analysis as a good alternative instead of CA. We must take into account that both techniques are closely related as they both are based on the same assumption, that is, reducing the data dimensions with a minimum information loss.

## 4. Conclusions

In this study we present a methodology for representing multivariate data in a low dimensional graph. Although many representation techniques have been applied in the field of scientometrics, emphasizing on analyzing their capability for representing with a minimum information loss multivariate data, Biplot analysis seems to be less known by this research community. We apply the JK-Biplot technique in three different case studies testing its efficiency in three different research evaluation contexts according to the aggregation levels (macro, meso and micro), different types of indicators (bibliometric and science indicators) and obtaining different results regarding the overall, row or column quality representation. We believe that, as well as it has been proved for other scientific fields, this methodology may well be an important analysis tool for bibliometric studies.

In this paper we focus on the Classical JK-Biplot analysis, however, other types of Biplot analysis should be studied in order to explore their possibilities and differences among each. We must especially mention the HJ-Biplot analysis as this type seems to overpass the limitations of the JK-Biplot analysis regarding the quality of representation for rows and columns. Although in this paper we have used small matrices for displaying the biplot analysis potential, we believe this type of analyses are of great interest and should be explored by the informetric research community, especially for studies regarding massive data sets for data mining (Theoharatos et al, 2007) and data classification patterns (Chapman et al., 2001). Finally, we must emphasize that, as well as other visual metaphors such as social networks analysis, this type of representations may be of great interest not just as research tools for analyzing variables, but also in the research policy arena as easy-to-read tools.

**Acknowledgments**

Thanks are due to the two anonymous reviewers for their constructive suggestions. Nicolás Robinson-García is currently supported by a FPU Grant from the Ministerio de Economía y Competitividad of the Spanish Government.

**References**

Alcántara, M., & Rivas, C. (2007). Las dimensiones de la polarización partidista en América Latina. Política y Gobierno, 14(2), 349-390.

Arias Díaz-Faes, A., Benito-García, N., Martín-Rodero, H., & Vicente-Villardón, J.L. (2011). Propuesta de aplicabilidad del método multivariante gráfico "BIPLOT" a los estudios bibliométricos en biomedicina. In XIV Jornadas Nacionales de Información y Documentación en Ciencias de la Salud, Cádiz, 13-15 de Abril de 2011. (Unpublished) [Conference Poster]. From: http://hdl.handle.net/10760/15998

Börner, K., Chen, C., & Boyack, K.W. (2003). Visualizing knowledge domains. Annual Review of Information Science and Technology, 37, 149-255.

Cárdenas, O., Galindo, P., & Vicente-Villar don, J.L. (2007). Los métodos Biplot: evolución y aplicaciones. Revista Venezolana de Análisis de Coyuntura, 13(1), 279-303.

Chapman, S., Shenk, P., Kazan, K., & Manners, J. (2001). Using Biplots to interpret gene expression pattern in plants. Bioinformatics Applications Note, 18(1), 202-204.

Gabriel, K.R. (1971). The Biplot graphic display of matrices with application to principal component analysis. Biometrika, 58(3), 453-467.

Gabriel, K.R. (2002). Goodness of fit of biplots and correspondence analysis. Biometrika, 89(2), 423-436.

Gabriel, K.R., & Odoroff, C.L. (1990). Biplots in biomedical research. Sta. Med., 9, 469-485.

Galindo, P. (1986). Una alternativa de representación simultánea: HJ Biplot. Qüestiió, 10(1), 13-23.

Galindo, P.V., Vaz, T.D., & Nijkamp, P. (2011).Institutional capacity to dynamically innovate: an application to the Portuguese case. Technological Forecasting and Social Change, 78(1), 3-12.

Groh, G., & Fuchs, C. (2011).Multi-modal social networks for modeling scientific fields. Scientometrics, 89(2), 569-590.

Klavans, R., & Boyack, K.W. (2009). Towards a consensus map of science. Journal of the American Society of Information Science and Technology, 60(3), 455-476.

Moya-Anegón, F., Herrero-Solana, V., & Jiménez-Contreras, J. (2006). A connectionist and multivariate approach to science maps: the SOM, clustering and MDS applied to library and information science research. Journal of Information Science, 32(1), 63-77.

Noyons, E. (2001). Bibliometric mapping of science in a science policy context. Scientometrics, 50(1), 83-98.

Noyons, E., & Calero-Medina, C. (2009). Applying bibliometric mapping in a high level science policy context. Scientometrics, 79(2), 473-490.

Pan, S., Chon, K., & Song, H.Y. (2008).Visualizing tourism trends: a combination of ATLAS.ti and Biplot. Journal of Travel Research, 46(3), 339-348.

Theoharatos, C., Laskaris, N.A., Economu, G., & Fotopoulos, S. (2007). On the perceptual organization of image databases using cognitive discriminative biplots. Eurasip Journal on Advances in Signal Processing, 68165.

Torres-Salinas, D., Delgado-López-Cózar, E., Moreno-Torres, J.G., & Herrera, F. (2011a). Rankings ISI de las universidades españolas por campos científicos: Descripción y resultados. El profesional de la información, 20(1), 111-122.

Torres-Salinas, D., Delgado-López-Cózar, E., Moreno-Torres, J.G., & Herrera, F. (2011b). A methodology for Institution-Field ranking based on a bidimensional analysis: the IFQ2A Index. Scientometrics, 88(3), 771-786.

Van Eck, N.J., Waltman, L., Dekker, R., & van den Berg, J. (2010).A comparison of two techniques for bibliometric mapping: multidimensional scaling and VOS. Journal of the American Society for Information Science and Technology, 61(12), 2405-2416.

Veiga de Cabo, J., & Martín-Rodero, H. (2011). Acceso Abierto: nuevos modelos de edición científica en entornos web 2.0. Salud Colectiva, 7(Supl. 1), S19-S27.

White, H. (2003). Pathfinder networks and author cocitation analysis: A remapping of paradigmatic information scientists. Journal of the American Society for Information Science and Technology, 54(5), 423-434.

White, H., & McCain, K. (1998). Visualization of literatures. Annual Review of Information Science of Science and Technology, 32, 99-169.

Wouters, L., Gohlmann, H.W., Bijnens, L., Kass, S.U., Molenberghs, G., & Lewi, P.J. (2003). Graphical exploration of gene expression data: a comparative study of three multivariate methods. Biometrics, 59(4), 1131-1139.

Wouters, P. (1999). The citation culture [Doctoral Thesis]. Amsterdam: Universiteit van Amsterdam. From: http://garfield.library.upenn.edu/wouters/wouters.pdf

Yan, W. K., Hunt, L.A., Sheng, Q.L., & Szlavnics, Z. (2000).Cultivar evaluation and mega-environment investigation based on the GGE Biplot. Crop Science, 40(3), 597-605.

**Appendix Biplot methodology in terms of spectral decomposition**

A Biplot is defined as a low-dimensional graph with a minimum loss of information of a given matrix of data $X_{(n \times p)}$, formed by markers $a_1, a_2, \ldots, a_n$ for rows and $b_1, b_2, \ldots, b_p$ for columns, chosen in such a way that each element $x_{ij}$, is an approximation to $x_{ij} = a_i^T b_j$ (Gabriel, 1971).

Markers $a_i$ for rows and markers $b_i$ for columns are represented in a space of a dimension $s \leq \rho$ where s is the number of axes and $\rho$ the range of X. Let $a_1, a_2, \ldots, a_n$ be markers for rows of matrix A and $b_1, b_2, \ldots, b_p$ markers for rows of matrix B, then:

$$X \cong AB'$$

where $\cong$ means that X approaches to the product from the right.

The structure of matrix X can then be visualized by representing the markers in a Euclidean space of s dimensions. When matrix X is of range 2 or 3, the representation can adjust perfectly to two or three dimensions; if not, we will need as many axes as the range of X. However, as mentioned above, a Biplot follows the same criterion as for factorial dimensional reduction techniques, therefore, only the two first axes are represented.

The markers are obtained firstly through Singular Value Decomposition (SVD) of matrix X and then, by factorizing the matrix as follows:

$A = U\Lambda^\gamma$ and $B = V\Lambda^{1-\gamma}$

where $0 \leq \gamma \leq 1$. Gabriel (1971) proposes different $\gamma$ to which he assigns different names. Two possible factorizations are:

$$X = A^0(B^*)' = A^*(B^0)'$$

Row Metric Preserving (JK Biplot): $A^* = U\Lambda$ and $B^0 = V$

Column Metric Preserving (GH Biplot): $A^0 = U$ and $B^* = V\Lambda$

Then, using the two or three first columns for factorizations of matrices A and B, we obtain biplots in two or three dimensions. Row Metric Preserving (RMP) and Column Metric Preserving (CMP) refer to the preservation of rows or columns' metrics during factorization.

Each factorization has a "principal factor" that emphasizes the singular values and a "standard factor" for which the singular values do not appear. In order to identifying them we use the (*) and (0) respectively.

When we use $\gamma = 1/2$ in the equations:

$A = U\Lambda^{1/2}$ and $B = V\Lambda^{1/2}$

we obtain a symmetric Biplot or SQRT Biplot where $AA = I$.

One of the most important aspects one must take into account when analyzing Biplot representations are the concepts of Quality of Representation (hereafter QR) which is referred to each row and column, and the Goodness of Fitness ($QR_{overall}$), which is defined as the cumulative qualities of representation for columns. Usually, a range of representation higher than two is used. Although a Biplot representation may have a high Goodness of Fit, this does not necessarily mean that a certain marker may be represented with a low QR. Regarding goodness of fit for variables and cases, Gabriel (2002) uses a function depending on the two first eigenvalues and the Biplot classification methodology used. In his case, he uses Correspondence Analysis and shows that such function is a good indicator for SQRT and only for GH and JK when values are close to 0.95.